\documentclass[conference]{IEEEtran}
\IEEEoverridecommandlockouts
\usepackage{cite}
\usepackage{amsmath,amssymb,amsfonts}
\usepackage{algorithmic}
\usepackage{algorithm}
\usepackage{graphicx}
\usepackage{textcomp}
\usepackage{xcolor}

\usepackage{multirow}
\usepackage{adjustbox}
\usepackage{makecell}
\usepackage{mathtools}
\usepackage{comment}
\usepackage[normalem]{ulem}
\usepackage[inline]{enumitem}
\usepackage{soul}
\usepackage{bm}
\usepackage{siunitx}
\usepackage{xcolor}
\usepackage{subfig}
\usepackage{amssymb}
\usepackage{tablefootnote}
\usepackage{braket}
\usepackage{algorithm}

\usepackage[font=small]{caption}
\newcolumntype{C}[1]{>{\centering\arraybackslash}m{#1}}
\newcolumntype{L}[1]{>{\raggedright\arraybackslash}m{#1}}

\def\BibTeX{{\rm B\kern-.05em{\sc i\kern-.025em b}\kern-.08em
    T\kern-.1667em\lower.7ex\hbox{E}\kern-.125emX}}
\begin{document}

\title{Efficient Entanglement Routing for Satellite-Aerial-Terrestrial Quantum Networks\\
}

\author{\IEEEauthorblockN{Yu Zhang\textsuperscript{*}, Yanmin Gong\textsuperscript{*}, Lei Fan\textsuperscript{\dag\ddag}, Yu Wang\textsuperscript{\S}, Zhu Han\textsuperscript{\dag}, and Yuanxiong Guo\textsuperscript{\P}}
\IEEEauthorblockA{\textsuperscript{*}Department of Electrical and Computer Engineering, University of Texas at San Antonio, San Antonio, TX, USA \\
\textsuperscript{\dag}Department of Electrical and Computer Engineering, University of Houston, Houston, TX, USA\\
\textsuperscript{\ddag}Department of Engineering Technology, University of Houston, Houston, TX, USA\\
\textsuperscript{\S}Department of Computer and Information Sciences, Temple University, Philadelphia, PA, USA \\
\textsuperscript{\P}Department of Information Systems and Cyber Security, University of Texas at San Antonio, San Antonio, TX, USA 
}
\thanks{The work is partially supported by the US NSF (Grant NO. CNS-2106761, CNS-2318663, CNS-2047761, CNS-2006604, CNS-2128378, ECCS-2302469, and CMMI-2222810), Toyota, Amazon, and Japan Science and Technology Agency (JST) Adopting Sustainable Partnerships for Innovative Research Ecosystem (ASPIRE) under JPMJAP2326.}
}

\maketitle

\begin{abstract}
In the era of 6G and beyond, space-aerial-terrestrial quantum networks (SATQNs) are poised to advance the development of a global-scale quantum Internet. These networks leverage free space optical satellite and aerial quantum networks to complement optical fiber-based terrestrial quantum networks to enable the distribution of high-fidelity quantum entanglement over long distances. However, establishing multi-hop end-to-end quantum entanglement remains highly challenging, not only due to time-varying link conditions and structural heterogeneity inherent in SATQNs{\color{black}, }but also because noise in quantum channels and imperfections in quantum operations can degrade the quality of entanglement. To address this challenge, we formulate an optimization problem that maximizes SATQN throughput by jointly optimizing routing path selection and entanglement generation rates (PS-EGR) while ensuring high entanglement fidelity. The resulting problem is a mixed-integer linear programming (MILP) formulation, which is {\color{black}NP-hard}. We propose a Benders' decomposition (BD)-based approach to solve this problem efficiently. Specifically, the MILP is decomposed into a master problem for binary routing path selection and a subproblem for continuous entanglement generation rate optimization. Numerical results validate the effectiveness of the proposed PS-EGR scheme, offering critical insights into the optimization and deployment of SATQNs.
\end{abstract}


\section{Introduction}
\label{sec: introduction}
With the advancement of quantum techniques, quantum networks hold immense potential for implementing revolutionary quantum applications, such as quantum teleportation (QT) \cite{ren2017ground, pirandola2015advances, ma2012quantum}, distributed quantum computing \cite{cacciapuoti2019quantum, cuomo2020towards, van2016path}, and quantum key distribution (QKD) \cite{scarani2009security, liao2017satellite, mehic2020quantum}. These applications rely on generating a long-distance quantum entanglement between end nodes in {\color{black}the} quantum network. Quantum entanglement is a phenomenon where two particles become interconnected, and therefore the state of one instantly affects the state of the other, regardless of distance \cite{chen2024redp}. In traditional terrestrial optical fiber connections, a primary challenge in distributing quantum entanglements over long distances is the exponential increase in photon loss with distance in optical fiber links. Consequently, to facilitate long-distance quantum entanglement distribution, quantum repeaters equipped with quantum memories are incorporated into the communication link. These repeaters perform entanglement swapping and purification \cite{dur1999quantum}, thereby ensuring high-quality end-to-end user entanglements. Extensive research has focused on improving terrestrial quantum networks \cite{10382435, pouryousef2023quantum, dai2020optimal, gu2024fendi, vardoyan2024bipartite, zeng2022multi}. For instance, Zeng \textit{et al.} \cite{zeng2022multi} addressed a terrestrial entanglement routing problem with the goal of maximizing both the number of quantum-user pairs and their expected throughput. To achieve this, they formulated the problem as two sequential integer programs, which were subsequently solved using a heuristic algorithm. However, solely relying on traditional terrestrial quantum networks poses significant challenges to realizing a global-scale quantum network. Enhancing terrestrial quantum communication quality in remote areas by adding more quantum repeaters can be economically prohibitive due to the complex natural environments and the extensive construction distances \cite{al2024characterizing}.

In recent years, integrating the terrestrial quantum network with the non-terrestrial quantum network has been regarded as an important technique to increase quantum network coverage and capacity \cite{zhang2022future}. Compared to optical fiber links in terrestrial quantum networks, free-space optical links in non-terrestrial quantum networks can substantially reduce both channel loss and decoherence. One major component of the non-terrestrial quantum networks is the satellite quantum network.  With orbital heights ranging from {\color{black}500~\si{km} to 35,786~\si{km},} the satellite quantum network can offer several advantages, including lower deployment costs and larger coverage compared with the terrestrial quantum network. For example, Yin \textit{et al.} \cite{yin2017satellite} successfully leveraged the Micius satellite to distribute quantum entanglement over {\color{black}1,200 \si{km}.} Li \textit{et al.} \cite{li2025microsatellite} utilized a microsatellite to conduct satellite-based quantum key distribution with multiple ground stations.  Huang \textit{et al.} \cite{huang2023socially} utilized satellite-based free space optical channels to establish inter-domain entangled paths for long-distance requests. They designed algorithms that plan inter-domain routing, balance domain loads, select optimal intra-domain paths, and choose trusted repeaters for teleporting data qubits. {\color{black}Wei \textit{et al.} \cite{wei2024entanglement} investigated the distribution of entanglement from satellites to ground stations to complement terrestrial quantum networks.}
\begin{figure*}[t]
\centering
    \subfloat[\label{fig:swapping1}]{\centering{\includegraphics[width=0.26\textwidth]{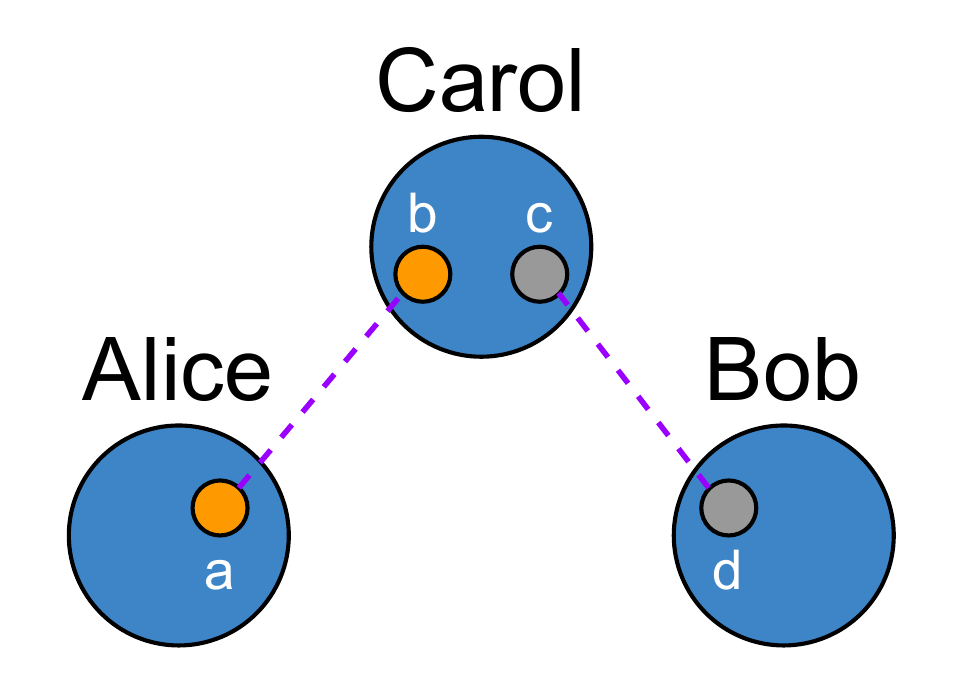} } }
    \hspace{1.5cm}
    \subfloat[ \label{fig:swapping2}]{\centering{\includegraphics[width=0.26\textwidth]{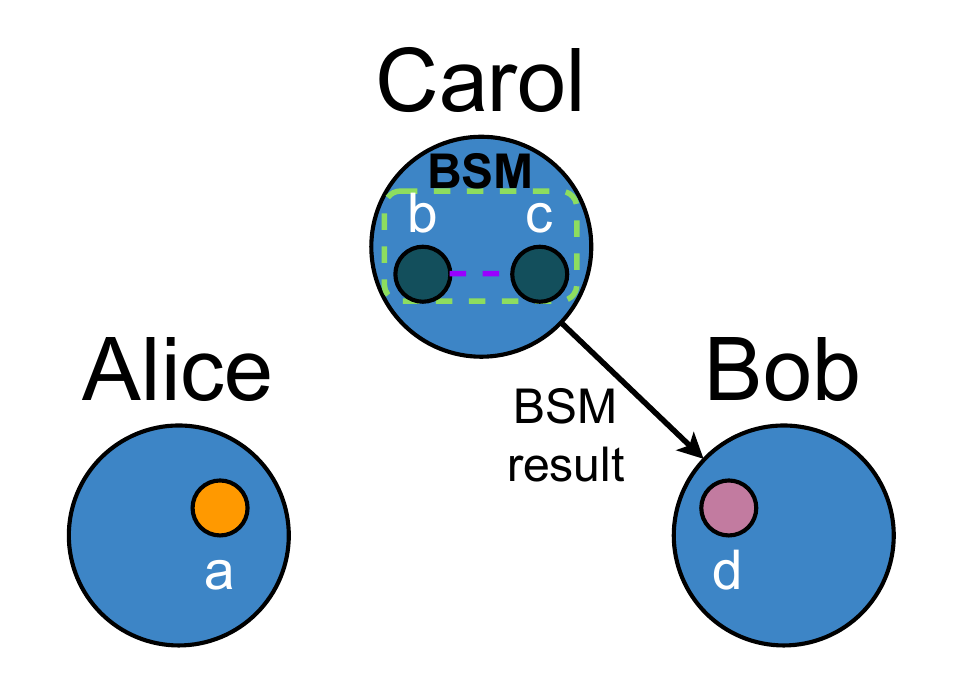} } }   
    \hspace{1.5cm}
    \subfloat[ \label{fig:swapping3}]{\centering{\includegraphics[width=0.26\textwidth]{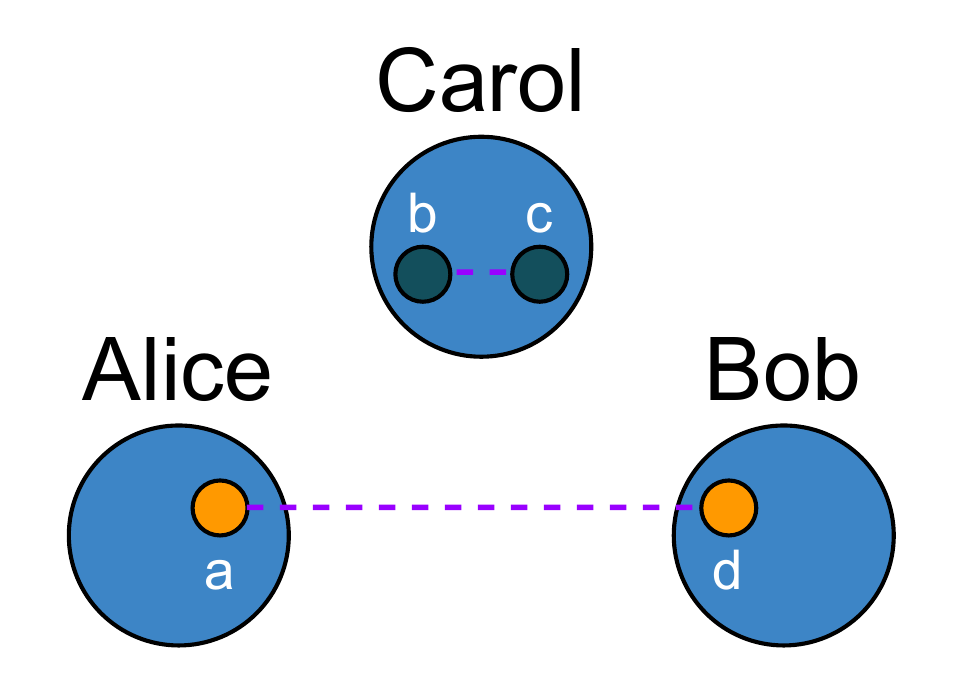} } }   
    \caption{ {\color{black}An illustration of entanglement swapping via quantum teleportation: a). Alice and Bob generate an EPR pair and send one qubit from their respective pairs to Carol, b). Carol performs a BSM and transmits the result to Bob, and c). Alice and Bob are entangled.}
    }
\label{fig:generation_and_swapping}%
\end{figure*}

Another key component of non-terrestrial quantum networks is the aerial quantum network that offers a promising wireless solution for complementing and enhancing the existing satellite and terrestrial quantum networks. Aerial quantum networks have fewer geographical restrictions than terrestrial quantum networks and provide a more flexible and cheaper network than satellite quantum networks. Liu \textit{et al.} \cite{liu2020drone} demonstrated the mobile entanglement distribution using unmanned aerial vehicles (UAVs), successfully operating in diverse weather conditions, including daytime and rainy nights. By integrating both satellite and aerial quantum networks with terrestrial quantum networks, the satellite-aerial-terrestrial quantum network (SATQN) offers a comprehensive and versatile solution to tackle diverse communication challenges and paves the way for the upcoming 6G network and beyond \cite{wang2023quantum, shaban2024sparq, khan2025integrated}. Wang \textit{et al.} \cite{wang2023quantum} explored the application of SATQNs to enhance federated learning. Shaban \textit{et al.} \cite{shaban2024sparq} utilized deep reinforcement learning (RL) to find the optimal path with the goal of achieving the highest entanglement quality in SATQNs. Khan \textit{et al.} \cite{khan2025integrated} investigated SATQNs for quantum anonymous communication (QAC) and developed key protocols to support SATQN-based QAC, including anonymous conference key agreement (CKA) and anonymous broadcast protocols.

Despite the aforementioned advantages, SATQNs face several critical challenges. Specifically, SATQNs inherit the inherent complexities of {\color{black} satellite-aerial-terrestrial integrated networks (SATINs) \cite{zhang2024quantum1, zhang2024quantum2, zhang2024quantum3}.} In contrast to the relatively static topology of terrestrial quantum networks, the topology of SATQNs is highly dynamic and comprises a large number of nodes, often in the order of thousands. This dynamic and large-scale structure introduces significant complexity, necessitating efficient coordination among heterogeneous network components to satisfy the entanglement demands of multiple ground users \cite{chiti2024survey}. Furthermore, the presence of noise in quantum channels and imperfections in quantum operations can lead to degradation in the quality of quantum entanglement, which is typically quantified by the end-to-end entanglement fidelity. Many quantum applications—such as QKD and distributed quantum computing—require a minimum fidelity threshold to ensure reliable performance.  To meet these fidelity requirements, entanglement purification operations can be applied to the established end-to-end entanglements. However, purification consumes a portion of the entangled qubits pairs, thereby reducing the SATQN throughput \cite{dur1999quantum}.  

To address these challenges,  {\color{black}we formulate an entanglement routing optimization problem that aims to maximize SATQN throughput by jointly optimizing routing path selection and the quantum entanglement generation rate (PS-EGR) while ensuring high entanglement fidelity to meet the requirements of end-user quantum applications.} The formulated problem is a mixed-integer linear program (MILP), which is NP-hard and computationally intractable using traditional brute-force search techniques \cite{floudas1995nonlinear}. To tackle this problem, we propose a Benders' decomposition (BD)-based algorithm. Specifically, the original MILP is decomposed into a master problem involving binary decision variables and a subproblem involving continuous decision variables \cite{rahmaniani2017benders}. Solving the master problem yields a lower bound on the optimal objective value, while solving the subproblem provides an upper bound. These two problems are then iteratively solved until their solutions converge. 

Our main contributions are summarized as follows.
\begin{itemize}
    \item We propose an advanced SATQN system in which multiple satellites, aerial platforms, and terrestrial base stations collaboratively provide quantum communication services to ground users.
    \item An entanglement routing optimization problem is formulated to jointly optimize path selection and quantum entanglement generation rates, with the objective of maximizing SATQN throughput while ensuring high entanglement fidelity.
    \item {\color{black}As the formulated problem is an intractable MILP, a BD-based method is employed to obtain the optimal solution. The problem is first decomposed into a master problem involving binary variables and a subproblem involving continuous variables. Then, they are solved iteratively until the upper and lower bounds converge.}
    \item Experimental results validate the effectiveness of the proposed scheme, demonstrating its performance improvements over multiple baselines.
\end{itemize}

{\color{black}This paper is organized as follows.} We first introduced the quantum network background in Section \ref{sec: QED}. Next, we introduced the SATQN system model and formulated an entanglement routing problem, focusing on maximizing quantum network throughput by jointly optimizing routing path selection and the quantum entanglement generation rate in Section \ref{sec: SBQN}. Following this, we designed a BD-based algorithm to efficiently solve the problem in {\color{black}Section~\ref{sec: HQCC}.}  Finally, we concluded the paper in Section \ref{sec: conclusion}.


\section{Quantum Network Background}
\label{sec: QED}
Unlike the classical Internet, which encodes information using binary bits (0's and 1's), quantum networks and quantum computing utilize quantum bits (qubits) as the fundamental units of quantum information. A qubit can be an elementary particle, such as an electron, photon, or atomic nucleus. A defining characteristic of a qubit is its ability to exist in a state representing both 0 and 1 simultaneously, a phenomenon known as superposition. Particularly, a qubit is usually represented as $\ket{\phi}=\alpha\ket{0} + \beta\ket{1}$, where $\alpha$ and $\beta$ are complex coefficients. Additionally, a group of qubits can form a highly correlated state that cannot be fully described by the individual states of the qubits alone—a phenomenon referred to as entanglement \cite{horodecki2009quantum}. For example, if two qubits $A$ and $B$ are maximally entangled, their state (i.e., Bell pairs) can be in one of the following four bases: $\Phi_{AB}^{\pm}=\frac{1}{\sqrt{2}}(\ket{0_A0_B}\pm\ket{1_A1_B})$ and $\Psi_{AB}^{\pm}=\frac{1}{\sqrt{2}}(\ket{0_A1_B}\pm\ket{1_A0_B})$. In this work, we refer to the Bell pair as the Einstein-Podolsky-Rosen (EPR) pair. When an EPR pair is shared between two users, secret quantum information can be transmitted from one to the other not through a physical link but via quantum measurement which is known as quantum teleportation. 

In quantum communication, entanglement generation, teleportation, and swapping facilitate the transfer of qubits between two arbitrarily distant nodes by utilizing a chain of intermediate nodes that perform a series of measurements and operations. For example, as illustrated in Fig. \ref{fig:generation_and_swapping}, Alice wants to transmit the quantum state of qubit $a$ to Bob, who is far away from her, with the assistance of a third party, Carol. Specifically, Fig. \subref*{fig:swapping1} depicts the initial step, in which Alice and Bob each generate an EPR pair, $(a, b)$ and $(c, d)$, respectively, and send one qubit from their respective pairs, namely $b$ and $c$, to Carol via quantum channels. Then, Carol performs a Bell-state measurement (BSM) on qubits $b$ and $c$, probabilistically projecting them onto one of the four Bell states \cite{ekert1991quantum}, as illustrated in Fig. \subref*{fig:swapping2}. The measurement result yields a 2-bit classical encoding corresponding to one of these Bell states. Carol then transmits the 2-bit result to Bob via classical communication. Based on this information, Bob applies up to two quantum gate operations—Pauli-Z and Pauli-X—to qubit $d$ \cite{ekert1991quantum}. As depicted in Fig. \subref*{fig:swapping2}, after teleportation, qubit $d$ acquires the same quantum state as qubit $b$ before the process and becomes entangled with qubit $a$, which is held by Alice (i.e., Fig. \subref*{fig:swapping3}). This whole process is also known as entanglement swapping.

In quantum communication, fidelity serves as a key metric for assessing the quality of entanglement. Due to environmental noise and imperfections in quantum operations, the fidelity of an EPR pair typically degrades over time. Low-fidelity EPR pairs can degrade the performance of various quantum applications \cite{cacciapuoti2019quantum}. This issue can be mitigated through entanglement purification \cite{wang2023efficient}, a process that typically consumes two low-fidelity EPR pairs to produce a single pair with higher fidelity. Since purification is probabilistic and both base pairs are discarded upon failure, the process can be repeated iteratively, purifying two newly generated EPR pairs until the target fidelity is achieved.

\section{System Model and Problem Formulation}
\label{sec: SBQN}
\begin{figure}[t]
\centering
\includegraphics[scale=0.3]{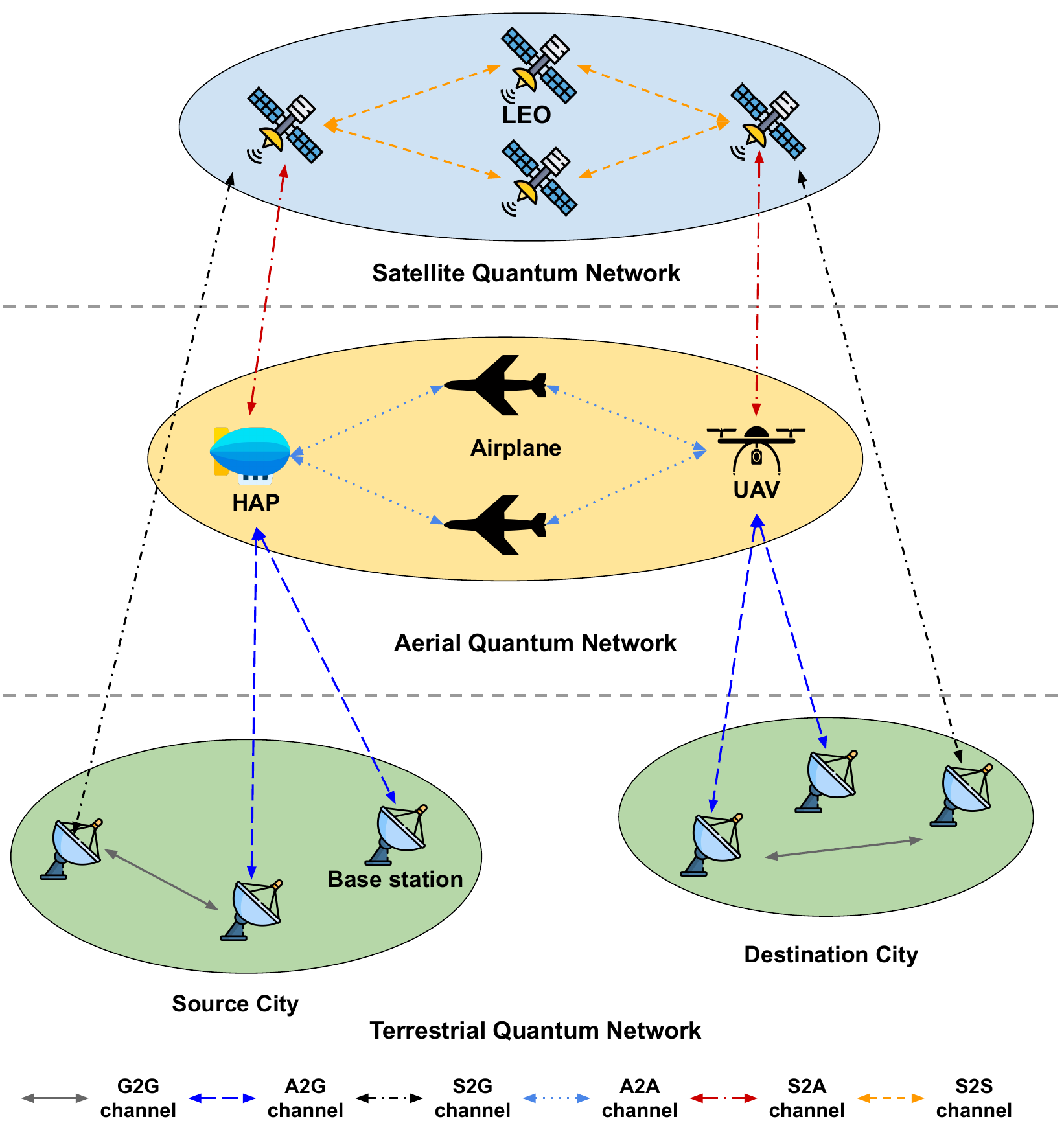}
\caption{An overview of SATQNs.}
\label{fig:system_model}
\end{figure}
\subsection{Satellite-Aerial-Terrestrial Quantum Networks}
As illustrated in Fig. \ref{fig:system_model}, we consider a general SATQN, which consists of a set of satellites $\mathcal{S}$, a set of aerial platforms $\mathcal{A}$, and a set of terrestrial base stations $\mathcal{B}$. There are multiple quantum channels in the SATQN, including satellite-to-satellite (S2S) channels, satellite-to-aerial (S2A) channels, aerial-to-aerial (A2A) channels, satellite-to-ground (S2G) channels, aerial-to-ground (A2G) channels, and ground-to-ground (G2G) channels. In the SATQN, we suppose there are $K$ source-destination pairs of terrestrial base stations, which is denoted as a set $k \in \mathcal{K} = \{1, \dots, K\}$. {\color{black}Each terrestrial base station may also be referred to as a ground user.} Each pair $k$ is represented by a tuple $k = (Q_{s(k)}, Q_{d(k)})$, where $Q_{s(k)}$ and $Q_{d(k)}$ denote the source and destination terrestrial base stations, respectively. Specifically, the source and destination terrestrial base stations are connected to SATQN and request quantum entanglement for applications such as QT and QKD. This process involves entanglement generation and routing path selection, both of which are coordinated by a central network controller equipped with a high-performance server. To describe the time-varying nature of this SATQN, the time horizon $T$ is divided into discrete time slots with slot duration $\delta$, indexed by $t\in \mathcal{T}=\{{1, \dots, T}\}$. The network topology of SATQN is assumed to be fixed during the time $t$, but may vary across different time slots. For simplicity, The SATQN can be further presented by a graph $\mathcal{G} = (\mathcal{V}, \mathcal{E})$, where the set of nodes is given by $\mathcal{V}=\mathcal{V}_s \cup \mathcal{V}_a \cup \mathcal{V}_b$. Here, $\mathcal{V}_s=\{v_i^t | i \in \mathcal{S}, t \in \mathcal{T}\}$, $\mathcal{V}_a=\{v_i^t | i \in \mathcal{A}, t \in \mathcal{T}\}$, and $\mathcal{V}_b=\{v_i^t | i \in \mathcal{B}, t \in \mathcal{T}\}$ represent the replicas of the set of satellite nodes, aerial platform nodes, and terrestrial base station nodes in the SATQN, respectively. The links $\mathcal{E}=\{ (v_i^t, v_j^t) | v_i^t, v_j^t \in \mathcal{V}, v_i \neq v_j, t \in \mathcal{T}\}$ indicate the transmission channel from nodes $v_i^t$ to $v_j^t$. Moreover, we assume a subset of the nodes $\mathcal{M} \subset \mathcal{V}$ is equipped with quantum memory, enabling them to store EPR pairs for satisfying future entanglement requests. The storage capacity of node $m \in \mathcal{M}$ is denoted by $C_m$. Table \ref{table: notation} summarizes the key notations used in this work.

\begin{table}[t!] 

\caption{List of Notations.}
\label{table: notation}
\centering
\begin{adjustbox}{width=\columnwidth,center}
\begin{tabular}{|C{2.2cm}|L{6.5cm}|}
\hline
\textbf{Notation} & \multicolumn{1}{l|}{\textbf{{\color{black}Definition}}}  \\
\hline
$\mathcal{S}$; $\mathcal{A}$; $\mathcal{B}$;  $\mathcal{K}$; $\mathcal{T}$& set of satellites; set of aerial platforms; set of terrestrial base stations; set of source-destination pairs; set of terrestrial base stations; set of time slots\\
\hline
$\mathcal{E}^{\text{S2S}}$; $\mathcal{E}^{\text{S2A}}$;  $\mathcal{E}^{\text{A2A}}$; $\mathcal{E}^{\text{S2G}}$; $\mathcal{E}^{\text{A2G}}$; $\mathcal{E}^{\text{G2G}}$ & S2S channel; S2A channel; A2A channel; S2G channel; A2G channel; S2G channel \\
\hline
$k$ & $k$-th source-destination terrestrial base station pair\\
\hline
$Q_{s(k)}$; $Q_{d(k)}$ & source terrestrial base station; destination terrestrial base station \\
\hline
$\mathcal{E}^{\text{FOC}}$; $\mathcal{E}^{\text{FSC}}$ & set of fiber optical channels; set of free space optical channels\\
\hline
$\mathcal{V}$; $\mathcal{V}_s$; $\mathcal{V}_a$; $\mathcal{V}_b$ & set of nodes; set of satellite nodes; set of aerial platform nodes; set of terrestrial base station nodes\\
\hline
$\mathcal{E}$; $\mathcal{M}$; $B_m$ & set of links; set of storage nodes; set of aerial platform nodes; the storage capacity of node $m$\\
\hline
$\eta_{i,j}^t$; $L_{i,j}^t$ & the transmittance of channel $(v_i, v_j)$; the fiber optical channel length of channel $(v_i, v_j)$\\
\hline
$\eta_{i,j}^{\text{lt}, t}$; $\eta_{i,j}^{\text{eff}, t}$; $\eta_{i,j}^{\text{atm}, t}$ & the turbulence-induced transmittance; the receiver’s efficiency; the atmospheric loss\\
\hline
$\eta_{i,j}^{\text{lt}, t}$; $\eta_{i,j}^{\text{eff}, t}$; $\eta_{i,j}^{\text{atm}, t}$ & the turbulence-induced transmittance; the receiver’s efficiency; the atmospheric loss\\
\hline
$a_\text{R}$; $w_{\text{lt}, i, j}$; $w_{\text{z},i, j}^t$ & the aperture radius of the receiver; the enlargement of the beam’s spot size\\
\hline
$w_0$; $R_0$; $h_{i,j}^t$ & the initial field spot size of the Gaussian beam; the radius of curvature; the free space optical distance between node $v_i$ and $v_j$\\
\hline
$\lambda$; $z_{\text{R}}$; $\zeta$ & the carrier wavelength; the beam’s Rayleigh length; the wave number\\
\hline
$C_n$; $l_0$; $\chi_0$ &  the refractive index structure constant; the inner scale of atmospheric turbulence; the altitude\\
\hline
$g(F^{\text{E2E}}_{p}, F^{\text{th}}_k)$ &  the average number of base EPR pairs required to generate one high-fidelity EPR pair \\
\hline
$F^{\text{E2E}}_p$; $\psi_u$ & the end-to-end fidelity of the source-destination pair $k$ along path $p$;  the number of successful purification steps required to achieve the target fidelity\\
\hline
$\mathcal{N}$; $\mathcal{P}_n^{\text{H}}$; $\mathcal{P}_k^{\text{H}}$ & the set of storage
pairs; the set of predefined paths between storage pair $n$; the set of predefined paths between source-destination pair $k$\\
\hline
$\mathcal{P}_n^{\text{U}}$; $\mathcal{P}_k^{\text{U}}$ & the set of virtual paths between storage pair $n$; the set of virtual paths between source-destination pair $k$\\
\hline
$Y_{k,p}^{t, \text{Max}}$; $C^t_{i,j}$ & the maximum EGR capacity of path $p$ for
source destination pair $k$; the capacity of link $(v_i, v_j)$\\
\hline
$x_{k,p}^t$; $y_{k,p}^t$; $z_{n,p}^t
$ & the path selection decision variable; the EGR decision variable; the average number of storage EPR pairs decision variable \\
\hline
\end{tabular}
\end{adjustbox}
\end{table}



\subsection{Loss and Noise Model}
In the SATQN, nodes $v_i$ and $v_k$ first each generate an EPR pair and send one qubit from their corresponding pairs to node $v_j$ via quantum channels. The quantum channel can be categorized into two types: {\color{black}1). }fiber optical channels $\mathcal{E}^{\text{FOC}} = \{ (v_i^t, v_j^t) | v_i^t, v_j^t \in \mathcal{V}_s, v_i \neq v_j, t \in \mathcal{T}\}$, and {\color{black}2).} free space optical channels $\mathcal{E}^{\text{FSC}} = \mathcal{E} \setminus \mathcal{E}^{\text{FOC}}$. Then, node $v_j$ conducts entanglement swapping.

\subsubsection{Fiber Optical Transmittance}Similar to previous works \cite{10382435, takeoka2014fundamental}, the channel between terrestrial base stations $v_i^t$ and $v_j^t$ is modeled as a fiber optical channel, with transmittance given by 
\begin{equation}
    \eta^{t}_{i,j}= e^{-\alpha_{i,j} L_{i,j}^t}, \quad \forall (v_i^t, v_j^t) \in \mathcal{E}^{\text{FOC}}, t \in \mathcal{T},
\end{equation}
where $\alpha$ is a constant depending on the physical media and $L$ is the physical length of the channel.

\subsubsection{Free Space Optical Transmittance} We consider the free-space optical quantum channel operating in the regime of moderate-to-strong turbulence, where the effects of beam widening and breaking dominate over beam wandering. According to \cite{ghalaii2022quantum}, the transmittance of free space optical channel between node $v_i^t$ and $v_j^t$ can be described as 
\begin{equation}
    \eta^{t}_{i,j}=\eta_{i,j}^{\text{lt}, t}\eta_{i,j}^{\text{eff}, t}\eta_{i,j}^{\text{atm}, t} , \quad \forall (v_i^t, v_j^t) \in \mathcal{E}^{\text{FSC}}, t \in \mathcal{T},
\end{equation}
where $\eta_{i,j}^{\text{lt}, t}$ is the turbulence-induced transmittance, $\eta_{i,j}^{\text{eff}, t}$ is the receiver’s efficiency, and $\eta_{i,j}^{\text{atm}, t}$ is the atmospheric loss. 

For a Gaussian beam, the turbulence-induced transmittance $\eta_{i,j}^{\text{lt}, t}$ between node $v_i^t$ and $v_j^t$ can be presented by 
\begin{equation}
    \eta_{i,j}^{\text{lt}, t}= 1 - e^{\frac{-2a_\text{R}^2}{w_{\text{lt}, i, j}^2}} , \quad \forall (v_i^t, v_j^t) \in \mathcal{E}^{\text{FSC}}, t \in \mathcal{T},
\end{equation}
where $a_\text{R}$ is the aperture radius of the receiver, and $w_{\text{lt}, i, j}$ is the long-term beam waist, which is given by $w_{\text{lt}, i, j}^t= w_{\text{z},i, j}^t\sqrt{1+\frac{4}{3}q_{i,j}^t\Lambda_{i,j}^t}$. Here, the parameter $q_{i,j}^t$ and $\Lambda_{i,j}^t$ characterize the Gaussian beam propagating between nodes $v_i^t$ and $v_j^t$. $w_{\text{z},i, j}^t$ is the corresponding enlargement of the beam’s spot size, which is written as $ w_{\text{z},i, j}^t=  w_0^2[(1-h_{i,j}^t/R_0)^2+\left(h_{i,j}^t/z_{\text{R}}\right)^2]$, where $w_0$ is the initial field spot size of the Gaussian beam, $R_0$ is the radius of curvature, $h_{i,j}^t$ is the distance between node $v_i^t$ and $v_j^t$, and $z_{\text{R}}=\pi w_0^2/\lambda$ is the beam's Rayleigh length with $\lambda$ as the carrier wavelength. Additionally,  we assume there is a strong turbulent space. The parameter $\Lambda_{i,j}^t$ is defined as $\Lambda_{i,j}^t = 2h_{i,j}^t/[\zeta (w_{\text{z},i,j}^t)^2]$, where $\zeta  = \frac{2\pi}{\lambda}$ denotes the wave number. The parameter $q_{i,j}^t$ is given by $q_{i,j}^t = 0.74(\sigma_{\text{Ry},i,j}^t)^2 (Q_{\text{m},i,j}^t)^{\frac{1}{6}}$. Here, $(\sigma_{\text{Ry},i,j}^t)^2 = 1.23 C_n^2 \zeta ^{\frac{7}{6}} (h_{i,j}^t)^{\frac{11}{6}}$ with $C_n$ representing the refractive index structure constant, and $Q_{\text{m},i,j}^t = 35.05h_{i,j}^t/(\zeta  l_0^2)$, where $l_0$ is the inner scale of atmospheric turbulence. 

Lastly, the atmospheric loss $\eta_{i,j}^{\text{atm}, t}$ can be described as  
\begin{equation}
    \eta_{i,j}^{\text{atm}, t}=  \exp\{-\alpha(\lambda, \chi_0)h_{i,j}^t\}, \quad \forall (v_i^t, v_j^t) \in \mathcal{E}^{\text{FSC}}, t \in \mathcal{T},
\end{equation}
where $\alpha(\lambda, \chi_0)=\alpha_0(\lambda)e^{-\frac{\chi_0}{6600}}$. Here, $\chi_0$ is the altitude, and $\alpha_0(\lambda)$ is the extinction factor at sea level.

\subsubsection{Amplitude Damping Channel} Similar to the approaches in \cite{shaban2024sparq, grassl2018quantum}, we model both fiber-optical and free-space optical channels as amplitude damping channels to capture the degradation of entanglement quality. Specifically, the amplitude damping channel is characterized by the transmittance $\eta_{i,j}^t$ and represented using the following Kraus operators:   
\begin{align}
B_{0,i,j}^t = \begin{bmatrix} 1 & 0 \\ 0 & \sqrt{\eta^{t}_{i,j}} \end{bmatrix},
B_{1,i,j}^t = \begin{bmatrix} 0 & \sqrt{1 - \eta^{t}_{i,j}} \\ 0 & 0 \end{bmatrix},& \notag\\
\forall (v_i^t, v_j^t) \in \mathcal{E}, t \in \mathcal{T}.&
\end{align}
Then, the effect of the amplitude damping channel on the density matrix $\rho_{i,j}^t$ is expressed as
\begin{align}
\Tilde{\rho}_{i,j}^{t} = B_{0,i,j}^t \rho_{i,j}^t (B_{0,i,j}^t)^\dagger + B_{1,i,j}^t \rho_{i,j}^t (B_{1,i,j}^t)^\dagger,& \notag\\
\forall (v_i^t, v_j^t) \in \mathcal{E}, t \in \mathcal{T}.&
\end{align}
According to \cite{shaban2024sparq}, the fidelity of qubits generated along link $(v_i, v_j)$ can be written as
\begin{equation}
F_{i,j}^t = \biggl( \text{Tr} \Bigl( \sqrt{\sqrt{\Tilde{\rho}_{i,j}^{t}} |\psi\rangle \langle \psi| \sqrt{\Tilde{\rho}_{i,j}^{t}}} \Bigr) \biggr)^2, \forall (v_i^t, v_j^t) \in \mathcal{E}, t \in \mathcal{T},
\end{equation}
where $|\psi\rangle$ denotes the ideal entangled state, specifically the maximally entangled Bell state given by $\frac{\left( |00\rangle + |11\rangle \right)}{\sqrt{2}}$. For simplicity, we denote the link as $e=(v_i, v_j) \in \mathcal{E}$, and define the link-based fidelity $F_{e}^t:=F_{i,j}^t$.

\subsubsection{Imperfect Entanglement Swapping} In practice, the entanglement swapping operation is subject to noise and imperfections, leading to additional fidelity degradation. This loss arises from the limited reliability of BSM as well as from errors in the single-qubit and two-qubit operations involved. For example, consider a swap performed using two elementary qubits with fidelities $F_1$ and $F_2$ at node $k$, where the accuracy of the BSM and the probabilities of ideal single-qubit and two-qubit operations are denoted by $\alpha_k$, $o_{1,k}$, and $o_{2,k}$, respectively. Define $W_e := \frac{4F_e - 1}{3}$ and $W_k := o_{1,k} o_{2,k} \frac{4\alpha_k^2 - 1}{3}$ for link $e$ and node $k$, respectively. According to \cite{gu2024fendi}, the fidelity of the resulting entangled qubit after a successful swapping operation is given by
\begin{equation}
F^\prime=\frac{1}{4}(1 + 3 W_1W_2W_k).
\end{equation}
{\color{black}Then, the end-to-end fidelity of source-destination pair $k$ along path $p$, which consists of links $\{e_1, e_2, \dots, e_X\} \subseteq \mathcal{E}$ and corresponding nodes $\{k_1, k_2, \dots, k_{X+1}\} \subseteq \mathcal{V}$, where $X$ and $X+1$ denote the number of links and nodes, respectively, can be expressed as follows:} 
\begin{equation} \label{equation: e2e fidelity}
F^{\text{E2E}}_p = \frac{1}{4} \left( 1 + 3  \prod_{j=1}^{X} W_{e_j}\prod_{i=1}^{X+1} W_{k_i} \right). 
\end{equation}

\subsection{Purification Model}
According to Equation \eqref{equation: e2e fidelity}, the end-to-end fidelity of a base EPR pair associated with the source-destination pair $k$ can be computed along path $p$. However, due to environmental noise and imperfections in quantum operations, the fidelity of the base EPR pair may not meet the threshold required for quantum applications. To improve fidelity, a recurrence-based purification protocol is applied at the end-user node \cite{dur1999quantum}. The average number of base EPR pairs $g(F^{\text{E2E}}_{p}, F^{\text{th}}_k)$ required to generate one high-fidelity EPR pair is represented as 
\begin{equation}
g(F^{\text{E2E}}_{p}, F^{\text{th}}_k) = \prod_{u=1}^{u_{\max}} \frac{2}{\psi_u},
\end{equation}
where $u_{\max}$ denotes the number of successful purification steps required to achieve the target fidelity $F^{\text{th}}_k$, starting from an base fidelity $F^{\text{E2E}}_{p}$, and $\psi_u$ represents the success probability of the $u$-th purification step.

\begin{figure}[t]
\centering
    \subfloat[Original graph \label{fig:original_graph}]{\centering{\includegraphics[width=0.22\textwidth]{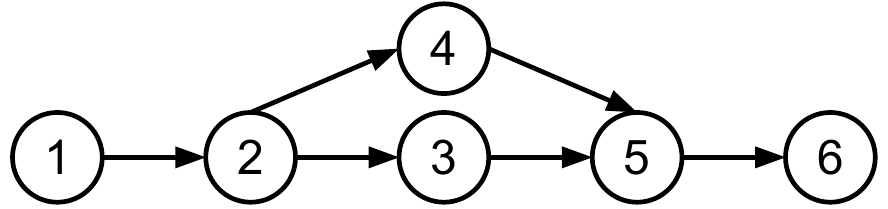} } }
    \hspace{0.1cm}
    \subfloat[Adding one virtual link for each path between storage nodes \label{fig:virtual_graph}]{\centering{\includegraphics[width=0.22\textwidth]{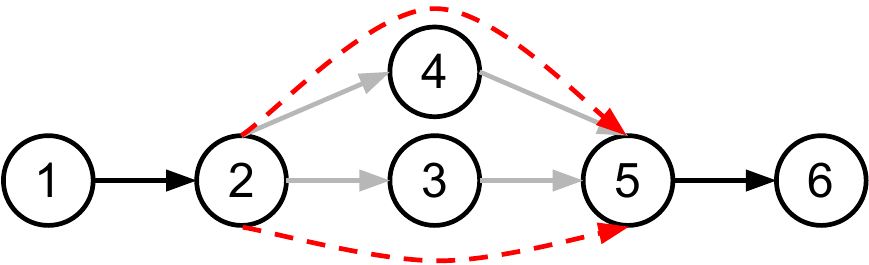} } }   
    \caption{An illustration of virtual paths. In (a), each node is connected to the others by a quantum channel, represented by a solid black line. In (b), parallel virtual links, represented by dashed red lines, are added between the storage pair (2, 5) to construct virtual paths for the user pair (1, 6).}
    \label{fig:virtual_path}%
\end{figure}
\subsection{Paths and Virtual Links}
In the SATQN, each source-destination pair $k$ has a set of predefined paths between them, which is denoted as $\mathcal{P}_k^{\text{H}}$. Similarly, two storage nodes $m_1, m_2 \in \mathcal{M}$ can form a storage pair, and the set of all such pairs is represented by $\mathcal{N}$. The set of paths associated with a storage pair $n = (m_1, m_2) \in \mathcal{N}$ is denoted by $\mathcal{P}_n^{\text{H}}$.

To efficiently represent the network structure and track the fidelities, we introduce virtual links, which represent parallel connections between each storage pair for each path that connects them \cite{pouryousef2023quantum}. Then, we define a virtual graph $\Tilde{\mathcal{G}}=\{\mathcal{V},\Tilde{\mathcal{E}}\}$ by adding virtual links between storage node pairs. A virtual path is defined as any path that includes at least one virtual link. As illustrated in Fig. \ref{fig:virtual_path}, we introduce two virtual links between the storage pair $(2,5)$ and construct two virtual paths $p_1$ and $p_2$ for the user pair $(1,6)$. Both paths are denoted by $\{1,2,5,6\}$, differ in their underlying sub-paths between nodes 2 and 5: $p_1$ utilizes sub-path $s_1 = \{2,4,5\}$, while $p_2$ utilizes sub-path $s_2 = \{2,3,5\}$. To clearly distinguish these two virtual paths, we define $f(\cdot)$ as a function that outputs the complete path corresponding to a virtual path. In this case, $f(p_1)=\{1,2,3,5,6\}$ and $f(p_2)=\{1,2,4,5,6\}$. Furthermore, we denote the set of all virtual paths associated with source-destination pair $k$ and storage pair $n$ as $\mathcal{P}_k^{\text{U}}$ and $\mathcal{P}_n^{\text{U}}$, respectively.

\subsection{SATQN Flow Model}
At time $t$, each source-destination pair $k$ selects a path $p$ from the combined set of actual paths and virtual paths, denoted by $\mathcal{P}_k=\mathcal{P}_k^{\text{H}}\cup \mathcal{P}_k^{\text{U}}$, according to the binary decision variable $x_{k,p}^t$. Here, $x_{k,p}^t = 1$ indicates that path $p \in \mathcal{P}_k$ is selected; otherwise, $x_{k,p}^t = 0$. Since the source-destination pair $k$ can select only one path at time $t$, we have
\begin{equation} \label{constr: path_selection}
\sum_{p \in \mathcal{P}_k}x_{k,p}^t = 1, \quad \forall k \in \mathcal{K}, t \in \mathcal{T}. 
\end{equation}

We further introduce a continuous variable $y_{k,p}^t$ to denote the entanglement generation rate (EGR) for each source-destination pair $k$ on path $p$ at time $t$. Note that EGR $y_{k,p}^t$ is only available when the corresponding path is selected. Therefore, we have  
\begin{equation} \label{constr: path_EGR}
y_{k,p}^t \leq x_{k,p}^tY_{k,p}^{t, \text{Max}} , 
\quad \forall k \in \mathcal{K}, p \in \mathcal{P}_k,t \in \mathcal{T}, 
\end{equation}
where $Y_{k,p}^{t, \text{Max}}$ is the maximum EGR capacity of path $p$ for source destination pair $k$. 

Let a continuous variable $z_{n,p}^t$ denote the average number of EPR pairs at storage pair $n \in \mathcal{N}$ on path $p$ at the beginning of time $t$. {\color{black}For simplicity, we define $\mathcal{P}_n = \mathcal{P}_n^{\text{H}} \cup \mathcal{P}_n^{\text{U}}$. Then, we have the following storage capacity constraint:
\begin{equation} \label{constr: storage}
\sum_{m_2 \in \mathcal{M}}\sum_{n=(m, m_2)\in \mathcal{N}}\sum_{p_m \in \mathcal{P}_n} z_{n,p_m}^t \leq B_m,  \quad  \forall  m \in \mathcal{M}, t \in \mathcal{T}.
\end{equation}}

To ensure that the EGR for serving or storing on link $(v_i,v_j)$ does not exceed its capacity $C^t_{i,j}$, {\color{black}we have the following constraint: 
\begin{align} \label{constr: link_capacity}
\sum_{k \in \{\mathcal{K}\cup\mathcal{N}\}} \sum_{ p \in \mathcal{P}_k| (v_i,v_j) \in p}y_{k,p}^tg(F^{\text{E2E}}_{p}, F^{\text{th}}_k) \leq C^t_{i,j}, &\notag \\
\forall t \in \mathcal{T}, (v_i,v_j) \in \mathcal{E}. &
\end{align}

Note that the average number of purified EPR pairs of serving from a storage pair at time $t$ does not exceed the available amount at the beginning of time $t$, which can be written as 
\begin{align} \label{constr: store_serve}
\sum_{k \in \{\mathcal{K}\cup \{\mathcal{N}-n\}\}}\sum_{p \in \mathcal{P}_k^{\text{U}}| p_m \subset f(p)}y_{k,p}^tg(F^{\text{E2E}}_{p}, F^{\text{th}}_k)\delta  \leq z_{n,p_m}^t&, \notag \\  
\forall t \in \mathcal{T}, n\in \mathcal{N}, p_m \in \mathcal{P}_n.&
\end{align}

Finally, for a given storage pair $n$, the sum of the number of EPR pairs at the storage in the current time slot and the pairs served to meet user demands and end-to-end purification in the previous time interval should equal the sum of the number of EPR pairs in storage during the previous time slot and the EPR pairs generated in the previous time interval, {\color{black}which can be represented as 
\begin{align} \label{constr: comprehensive}
   z_{n,p_m}^t + \sum_{k \in \{\mathcal{K}\cup \{\mathcal{N}-n\}\}}\sum_{p \in \mathcal{P}_k^{\text{U}}| p_m \subset f(p)}y_{k,p}^{t-1}g(F^{\text{E2E}}_{p}, F^{\text{th}}_k)\delta \notag\\  
   = z_{n,p_m}^{t-1} + y_{n,p_m}^{t-1}\delta, 
  \quad \forall t \in \mathcal{T}, n\in \mathcal{N}, p_m \in \mathcal{P}_n. 
\end{align}}

\subsection{Problem Formulation}
\label{sec: OEDP}
In this work, we investigate the entanglement routing problem to maximize quantum network throughput (i.e., the aggregate weighted EGR across all user pairs and all time intervals) by jointly optimizing path selection $\mathbf{x}=\{x_{k,p}^t\}$, EGR $\mathbf{y}=\{y_{k,p}^t\}$, and the average number of
EPR pairs storage $\mathbf{z}=\{z_{n,p}^t\}$, i.e.,  
\begin{subequations}
\label{P0}
\begin{align}
\textbf{P}_0: \max_{\mathbf{x,y,z}} \quad &  \sum_{t \in \mathcal{T}}\sum_{k \in \mathcal{K}}\sum_{p \in \mathcal{P}_k} \alpha_k^t y_{k,p}^t  \tag{\ref{P0}}\\
\text{s.t.}
\quad & \mathbf{x} \in \{0,1\}, \label{p0_a}\\
\quad & \mathbf{y,z} \geq 0, \label{p0_b}\\
\quad & \eqref{constr: path_selection}-\eqref{constr: comprehensive},
\end{align}
\end{subequations}
where $\alpha_{k}^t$ is the weight associated with source-destination pair $k$ at time interval $t$. By solving this problem, we can obtain the optimal path selection, EGR, and the average number of EPR pairs storage. However, as we can see, the continuous decision variables $\mathbf{y,z}$ are tightly coupled with binary decision variables $\mathbf{x}$, which makes problem $\textbf{P}_0$ a MILP that is challenging to solve.

\begin{align}
\textbf{P}_0: \max_{\mathbf{x,y,z}} \quad &  \sum_{t \in \mathcal{T}}\sum_{k \in \mathcal{K}}\sum_{p \in \mathcal{P}_k} \alpha_k^t y_{k,p}^t \\
\text{s.t.}
\quad & \sum_{p \in \mathcal{P}_k}x_{k,p}^t = 1, \quad \forall k \in \mathcal{K}, t \in \mathcal{T},\\
\quad & y_{k,p}^t \leq x_{k,p}^tY_{k,p}^{t, \text{Max}} , 
\quad \forall k \in \mathcal{K}, p \in \mathcal{P}_k,t \in \mathcal{T}, \\
\quad & \sum_{m_2 \in \mathcal{M}}\sum_{n=(m, m_2)\in \mathcal{N}}\sum_{p_m \in \mathcal{P}_n} z_{n,p_m}^t \leq B_m,  \notag \\
\quad &  \quad  \forall  m \in \mathcal{M}, t \in \mathcal{T},\\
\quad & \sum_{k \in \{\mathcal{K}\cup\mathcal{N}\}} \sum_{ p \in \mathcal{P}_k| (v_i,v_j) \in p}y_{k,p}^tg(F^{\text{E2E}}_{p}, F^{\text{th}}_k) \leq C^t_{i,j}, \notag \\
\quad & \quad
\forall t \in \mathcal{T}, (v_i,v_j) \in \mathcal{E}. \\
\quad & \sum_{k \in \{\mathcal{K}\cup \{\mathcal{N}-n\}\}}\sum_{p \in \mathcal{P}_k^{\text{U}}| p_m \subset f(p)}y_{k,p}^tg(F^{\text{E2E}}_{p}, F^{\text{th}}_k)\delta  \leq z_{n,p_m}^t, \notag \\
\quad & \quad  
\forall t \in \mathcal{T}, n\in \mathcal{N}, p_m \in \mathcal{P}_n, \\
\quad &    z_{n,p_m}^t + \sum_{k \in \{\mathcal{K}\cup \{\mathcal{N}-n\}\}}\sum_{p \in \mathcal{P}_k^{\text{U}}| p_m \subset f(p)}y_{k,p}^{t-1}g(F^{\text{E2E}}_{p}, F^{\text{th}}_k)\delta
   = z_{n,p_m}^{t-1}  \notag \\ 
 \quad & + y_{n,p_m}^{t-1}\delta, \quad \forall t \in \mathcal{T}, n\in \mathcal{N}, p_m \in \mathcal{P}_n,\\
\quad & \mathbf{x} \in \{0,1\}, \\
\quad & \mathbf{y,z} \geq 0\\
\end{align}

\section{{\color{black}Benders' Decomposition-Based Solution}}
\label{sec: HQCC}
{\color{black}
We develop a BD-based algorithm to solve the formulated MILP problem $\textbf{P}_0$ efficiently. Specifically, we first decompose problem $\textbf{P}_0$ into a master problem and a subproblem. The master problem only involves the binary optimization decision variables, while the subproblem only involves the continuous optimization decision variables. We can obtain a performance upper bound of problem $\textbf{P}_0$ by solving the subproblem and a performance lower bound of problem $\textbf{P}_0$ by solving the master problem. Then, problem $\textbf{P}_0$ is iteratively solved until the lower and upper bounds converge. In order to apply our algorithm, we first reformulate problem $\textbf{P}_0$ in an abstract formulation as
\begin{subequations} 
\label{P1}
\begin{align}
 \textbf{P}_1: \max_{\mathbf{x, q}} \quad & \mathbf{c}^\intercal\mathbf{q} \\
\text{s.t.}
\quad & \mathbf{H}_1\mathbf{q}+ \mathbf{G}_1\mathbf{x} \leq \mathbf{d}_1, \\
\quad & \mathbf{H}_2\mathbf{q}+ \mathbf{G}_2\mathbf{x} = \mathbf{d}_2, \\
\quad & \mathbf{G}_3\mathbf{x} = \mathbf{d}_3, \\
\quad & \mathbf{q} \geq 0,\\
\quad & \mathbf{x} \in \{0,1\},
\end{align}
\end{subequations}
where the symbol $\mathbf{q}=\left\{\mathbf{y,z}\right\}$, and the symbol $\mathbf{x}=\left\{\mathbf{x}\right\}$. $\mathbf{q}$ is a continuous variable vector, and $\{\mathbf{H}_n\}_{n = 1,2}$ is its corresponding coefficient matrix in constraints. $\mathbf{x}$ is a binary variable vector, and $\{\mathbf{G}_n\}_{n = 1,2,3}$ is its corresponding coefficient matrix in constraints. $\mathbf{c}^\intercal$ is the coefficient vector of variable $\mathbf{q}$ in the objective function. $\{\mathbf{d}_n\}_{n = 1,2,3}$ is the corresponding right-hand side constant vector. We describe the details of the subproblem and master problem in the following.

\subsubsection{Subproblem Optimization}
For the given binary variables $\mathbf{x}^{(l)}$ generated by the master problem at the $(l-1)$-th iteration, the subproblem can be written as
\begin{subequations} 
\begin{align}
 \textbf{SP}_1: \max_{\mathbf{q}} \quad &  \mathbf{c}^\intercal\mathbf{q} \\
\text{s.t.}
\quad & \mathbf{H}^*\mathbf{q} \leq \mathbf{d}^* - \mathbf{G}^*\mathbf{x}^{(l)}, \\
\quad & \mathbf{q} \geq 0,\\
\quad & \mathbf{H}^*=[\mathbf{H}_1, \mathbf{H}_2]^\intercal,\quad \mathbf{d}^*=[\mathbf{d}_1, \mathbf{d}_2]^\intercal, \\
\quad & \mathbf{G}^*=[\mathbf{G}_1, \mathbf{G}_2]^\intercal.
\end{align}
\end{subequations}
According to the duality theory of linear programming (LP), the dual problem of subproblem $\textbf{SP}_1$ can be expressed as
\begin{subequations} \label{sp_p1} 
\begin{align}
 \textbf{SP}_2: \min_{\mathbf{u}} \quad &   (\mathbf{d}^* -\mathbf{G}^*\mathbf{x}^{(l)})^\intercal \mathbf{u} \\
\text{s.t.}
\quad & ({\mathbf{H}^*})^\intercal\mathbf{u} \geq  \mathbf{c}, \label{sp_p1_1}\\
\quad & \mathbf{u} \geq  0, \label{sp_p1_2}\\
\quad & \mathbf{H}^*=[\mathbf{H}_1, \mathbf{H}_2]^\intercal, \quad
 \mathbf{d}^*=[\mathbf{d}_1, \mathbf{d}_2]^\intercal, \\
\quad & \mathbf{G}^*=[\mathbf{G}_1, \mathbf{G}_2]^\intercal,
\end{align}
\end{subequations}
where $\mathbf{u}$ is the dual variable. This problem can be efficiently solved by classical LP numerical solvers such as Gurobi \cite{Gurobi}. We also check the feasibility of the subproblem $\textbf{SP}_2$. Specifically, if the inner product between $ (\mathbf{d}^* -\mathbf{G}^*\mathbf{x}^{(l)})^\intercal$ and its extreme ray $\omega_1$ is negative, the subproblem is infeasible. Then, we can generate a new feasibility cut $C^{\text{F}}$, i.e., 
\begin{equation}
    (\mathbf{d}^* -\mathbf{G}^*\mathbf{x}^{(l)})^\intercal \omega_1 \geq 0.
\end{equation}
On the other hand, if the subproblem is feasible, we yield a new optimality cut $C^{\text{O}}$ by its extreme point $\omega_2$, i.e., 
\begin{equation}
    (\mathbf{d}^* -\mathbf{G}^*\mathbf{x}^{(l)})^\intercal \omega_2 \geq \theta,
\end{equation}
where $\theta$ is the optimal value of subproblem $\textbf{SP}_2$.  Both the feasibility cuts and optimality cuts generated in the current iteration are used to guide the future direction of the master problem in the next iteration.

\subsubsection{Master Problem Optimization} 
After solving subproblem $\textbf{SP}_2$, we can obtain a new extreme ray $\omega_1^{(l)}$ or point $\omega_2^{(l)}$ for feasibility cut or optimality cut at the $l$-th iteration, respectively. Moreover, we define $\Psi_1$ and $\Psi_2$ as the sets of all known iteration indices associated with the subproblem being infeasible and feasible, respectively. Then, we can formulate the master as 
\begin{subequations} 
\label{MP1}
\begin{align}
 \textbf{MP}_1: \max_{\mathbf{x}, \nu} \quad &  \nu \\
\text{s.t.}
\quad & \mathbf{G}_3\mathbf{x} = \mathbf{d}_3, \label{MP_1: original}\\
\quad &  (\mathbf{d}^* -\mathbf{G}^*\mathbf{x})^\intercal \omega_1^{(l)} \geq 0, \quad \forall l \in \Psi_1, \label{MP_1: feasiable}\\
\quad & (\mathbf{d}^* -\mathbf{G}^*\mathbf{x})^\intercal \omega_2^{(l)} \geq \nu , \quad \forall l \in \Psi_2, \label{MP_1: optimal}\\
\quad & \mathbf{x} \in \{0,1\}.
\end{align}
\end{subequations}
 By adding the feasibility cuts and optimality cuts, the search region for the globally optimal solution is gradually reduced \cite{rahmaniani2017benders}. Besides, the objective value of problem $\textbf{MP}_1$ is the performance lower bound of problem $\textbf{P}_1$ at the $l$-th iteration. 
 
\begin{algorithm}[t!]
\caption{The BD-based Algorithm} 
\label{alg:1}
\begin{algorithmic}[1]
\REQUIRE Iteration index $l = 1$, maximum iteration number $L^{\text{max}}$, iteration threshold $\epsilon$, $UB^{(0)} = +\infty$, and $LB^{(0)}=-\infty$. Initialize $\mathbf{x}^{(0)}$.
\WHILE{$|\frac{\text{UB}^{(l-1)}-\text{LB}^{(l-1)}}{\text{UB}^{(l-1)}}| > \epsilon$ or $l < L^{\text{max}}$}
        \STATE  $C^{\text{F}, (l)}, C^{\text{O}, (l)}, \text{UB}^{(l-1)}$ $\leftarrow \textbf{SUB($\mathbf{x}^{(l-1)}, \text{UB}^{(l-1)}$)}$. 
        \STATE Add $C^{\text{F}, (l)}$ or $C^{\text{O}, (l)}$ to the master problem $\textbf{MP}_1$ and update $\text{UB}^{(l)} \leftarrow \text{UB}^{(l-1)}$. \label{alg1:add_cut}
        \STATE Solve the the master problem $\textbf{MP}_2$. \label{alg1:solve_master}
        \STATE Obtain the optimal solution $\mathbf{x}^{(l)}$ and $\nu^{(l)}$.
        \STATE  Update $\text{LB}^{(l)} \leftarrow \nu^{(l)}$. \label{alg1:update_lb}
    \STATE Set $l \leftarrow l$ + 1.
\ENDWHILE 
\item[]
\item[]

 \hspace{-1.5em} \nonumber \textbf{SUB($\mathbf{x}^{(l-1)}, \text{UB}^{(l-1)})$}:
     \STATE Fix $\mathbf{x}$ as $\mathbf{x}^{(l-1)}$, and solve the subproblem $\textbf{SP}_2$. \label{alg1:sub}
     \IF{subproblem $\textbf{SP}_2$ is infeasible}
      \STATE Obtain the feasibility cut $C^{\text{F}, (l)}$. \label{alg1:feasible_cut}
      \ELSE
      \STATE Obtain the optimal solution $\mathbf{u}^{(l)}$, optimality cut $C^{\text{O}, (l)}$, and objective value $\theta^{(l)}$. \label{alg1:optimal_cut}
    \STATE $\text{UB}^{(l-1)} \leftarrow \min\{\text{UB}^{(l-1)}, \theta^{(l)}\}$.
     \ENDIF
\RETURN  $C^{\text{F}, (l)}$, $C^{\text{O}, (l)}$, and $\text{UB}^{(l-1)}$.
\item[]
\ENSURE Optimal $\mathbf{x}^*, \mathbf{q}^*$. 
\end{algorithmic}
\end{algorithm}

\subsubsection{Overall Algorithm}
The overall BD-based algorithm is summarized in Algorithm \ref{alg:1}. The algorithm contains an iterative procedure.  We initialize the binary variables $\mathbf{x}$ and other parameters at the beginning. In the $l$-th iteration, we first execute the SUB module. Specifically, we fix the binary variables $\mathbf{x}^{(l)}$ and solve the subproblem $\textbf{SP}_2$ (Line \ref{alg1:sub}). If the subproblem $\textbf{SP}_2$ is infeasible, we can obtain a new feasibility cut  $C^{\text{F}, (l)}$ (Line \ref{alg1:feasible_cut}). Otherwise, we obtain a new optimality cut $C^{\text{O}, (l)}$ and update the performance upper bound $\text{UB}^{(l-1)}$ by the optimal solution of the subproblem $\textbf{SP}_2$ (Line \ref{alg1:optimal_cut}). Then, we add the obtained feasibility cut $C^{\text{F}, (l)}$ or optimality cut $C^{\text{O}, (l)}$ to the master problem $\textbf{MP}_1$ (Line \ref{alg1:add_cut}). Then, we solve the master problem $\textbf{MP}_1$ and update the performance lower bound $\text{LB}^{(l)}$ (Lines \ref{alg1:solve_master}-\ref{alg1:update_lb}). This iteration procedure stops until the approximation gap $|\frac{\text{UB}^{(l)}-\text{LB}^{(l)}}{\text{UB}^{(l)}}|$ is within a preset threshold $\epsilon$ or the maximal iteration index $L^{\text{max}}$ is reached.}

\section{Numerical Experiments}
\label{sec: NE}
In this section, we evaluate the performance of the proposed optimal path selection and EGR (PS-EGR) scheme through numerical experiments. All experiments are implemented using Gurobi solver \cite{Gurobi} in Python 3.10.

\subsection{Simulation Setup}
We consider an SATQN consisting of 2 satellites, 2 high-altitude platforms (HAPs), and 12 terrestrial base stations. Specifically, 6 terrestrial base stations are located in the source city, with three initiating entanglement requests to their corresponding stations in the destination city, resulting in $K=3$ source-destination terrestrial base station pairs. The storage capacity of each storage node $B_m$ is set to $1500$ EPR pairs. Besides, we consider $10$ time slots, each with a duration of $20$ seconds. The threshold for fidelity requirement of all requests at all time intervals is set to 0.8. The rest of our simulation parameters, unless otherwise stated, are given in Table~\ref{talbe: simulation parameters}.

\begin{table}[t!] 
\caption{Simulation Parameters.}
\label{talbe: simulation parameters}
\centering
\begin{adjustbox}{width=1\columnwidth,center}
\begin{tabular}{|l|c|}
\hline
\multicolumn{1}{|l|}{\textbf{{\color{black}Parameter}}}  & \textbf{Value}  \\
\hline
satellite altitude & 500\si{km}\\
\hline
HAP altitude & 50\si{km} \\
\hline
receiver’s efficiency $\eta_{i,j}^{\text{eff}, t}$ & 1 \\ 
\hline
aperture radius of the receiver $\alpha_{\text{R}}$ & 5\si{m} \\
\hline
extinction factor at sea level $\alpha_0(\lambda)$ & $5 \times 10^{-6}$\si{m^{-1}} \\
\hline
inner scale of atmospheric turbulence $l_0$ & 1\si{mm}\\ 
\hline
initial field spot size of the Gaussian beam $w_0$ & 20\si{cm}  \\
\hline
carrier wavelength $\lambda$ & 800\si{nm}  \\
\hline
\end{tabular}
\end{adjustbox}
\end{table}

We evaluate the performance of PS-EGR against the following three baselines: 
\begin{enumerate}
\item Satellite-Only Scheme (SOS): All entanglement requests are managed by satellites without routing through terrestrial base stations or HAPs.
\item Random Routing Scheme (RRS): The EGR is optimized, while the routing path is randomly selected without any optimization.
\item No-Storage Scheme (NSS): None of the nodes has storage capacity.
\end{enumerate}

\begin{figure}[t]
\centering
\includegraphics[scale=0.5]{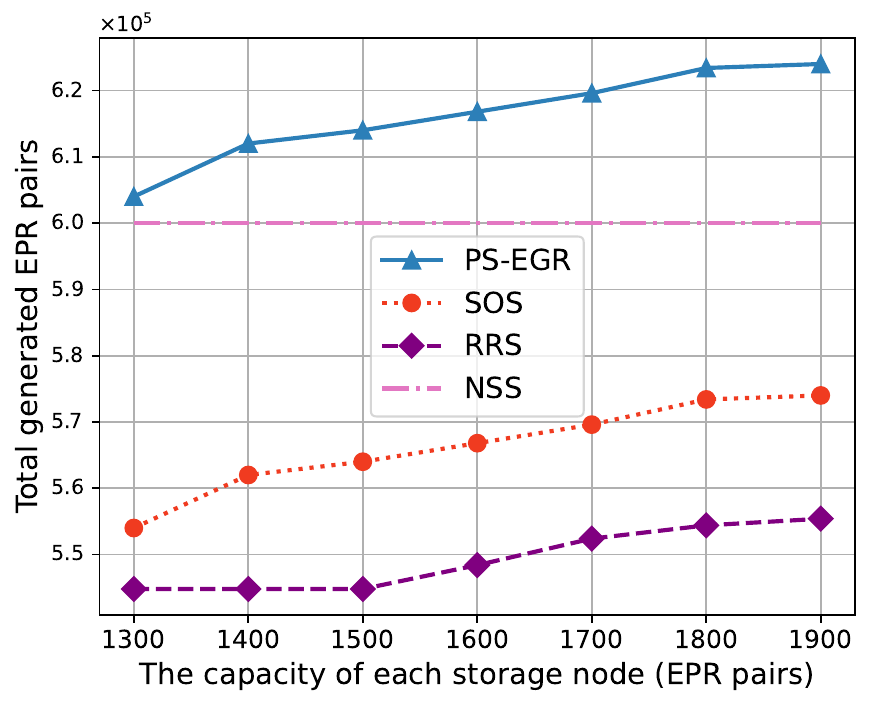}
\caption{The impact of storage node's capacity on network performance.}
\label{fig:impact_storage_node}
\end{figure}
\subsection{{\color{black}Impact of Storage Node's Capacity $\bm{B_m}$}}
In this subsection, we study how the storage node's capacity $\bm{B_m}$ affects the system performance in terms of the quantum network throughput (i.e., total generated EPR pairs). To that end, we increase the storage node's capacity $\bm{B_m}$ from $1,300$ to $1,900$ EPR pairs. From Fig.~\ref{fig:impact_storage_node}, we can observe that our PS-EGR scheme achieves the highest total generated EPR pairs compared to the baselines across various sizes of storage node's capacity. Moreover, the total generated EPR pairs of all schemes increase as the storage node's capacity increases, except that the total generated EPR pairs of the NSS scheme remain stable. This trend highlights that increased storage capacity enables better data retention until transmission to the next suitable node. Compared to RRS, a larger storage capacity significantly enhances network performance for the proposed PS-EGR. This improvement is due to PS-EGR's greater flexibility in selecting the entanglement routing path, thereby fully utilizing the storage node's capacity.

\begin{figure}[t]
\centering
\includegraphics[scale=0.5]{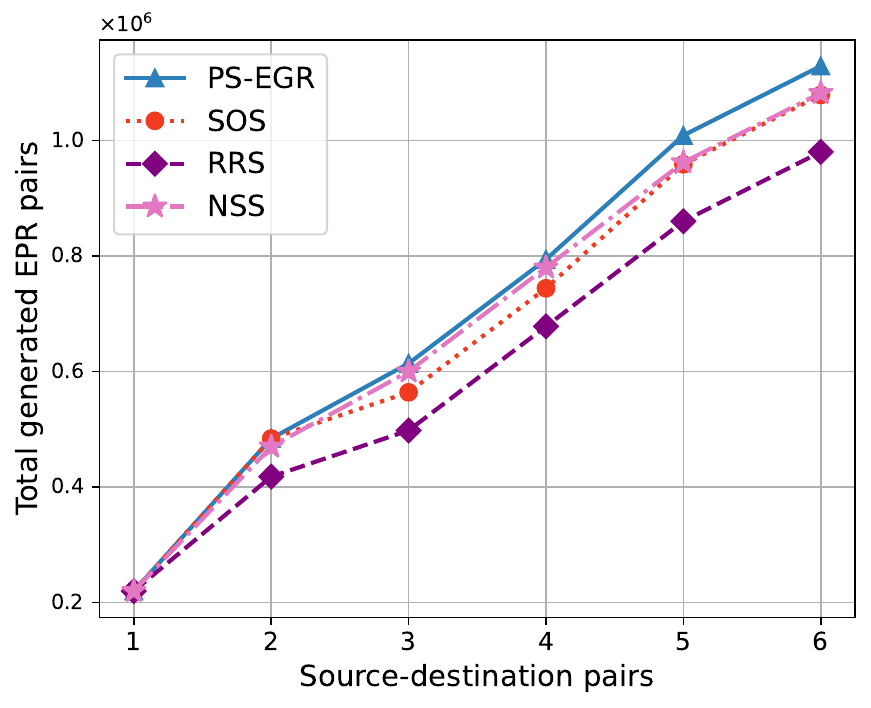}
\caption{The impact of source-destination pairs on network performance.}
\label{fig:impact_SD}
\end{figure}

\subsection{\color{black}{Impact of Source-destination Pairs $\bm{K}$}}

In this part, we compare the total generated EPR pairs of our proposed PS-EGR scheme with the baseline schemes regarding the number of source-destination pairs. To that end, we increase the number of source-destination pairs from 1 to 6. As shown in Fig.~\ref{fig:impact_SD}, the total generated EPR pairs increase across all schemes with the increasing number of source-destination pairs. The reason is that higher user demand leads to more entanglement requests and, consequently, greater quantum network throughput. Notably, the PS-EGR scheme consistently outperforms the baselines, highlighting its superior efficiency. The reason is that the proposed scheme jointly optimizes the routing path selection and EGR, which enables more effective support for ground users.

\section{Conclusion} 
\label{sec: conclusion}

{\color{black}In this paper, we proposed a novel SATQN architecture in which satellites, aerial platforms, and terrestrial base stations collaboratively provide quantum communication services to ground users. To maximize quantum network throughput, we formulated an entanglement routing problem that jointly optimizes path selection and EGR. To address the computational complexity of the MILP problem, we developed a BD-based algorithm that efficiently decomposes it into a master problem and a subproblem, which are solved iteratively. Extensive simulations demonstrate the superiority of the proposed PS-EGR scheme over various baseline methods. This study not only advances the understanding of entanglement routing in SATQNs but also underscores the importance of integrating satellite, aerial, and terrestrial components to enable next-generation quantum communication infrastructures for 6G and beyond.}

\bibliographystyle{IEEEtran}
\bibliography{IEEEabrv,ref}

\end{document}